\documentclass[]{article}
\usepackage{amsmath}
\usepackage{graphicx}
\usepackage{amssymb}
\usepackage{makeidx}
\usepackage{setspace}
\usepackage{color}
\usepackage{epsfig}
\usepackage{rotating}
\usepackage{graphicx}
\usepackage{dcolumn}% Align table columns on decimal point
\usepackage{bm}% bold math
\usepackage{color}
\usepackage{lscape}
\makeatother
\newcommand{\dm}{\begin{displaymath}}
\newcommand{\edm}{\end{displaymath}}
\newcommand{\beq}{\begin{equation}}
\newcommand{\eeq}{\end{equation}}
\newcommand{\beqa}{\begin{eqnarray}}
\newcommand{\eeqa}{\end{eqnarray}}
\newcommand{\ba}{\begin{array}}
\newcommand{\ea}{\end{array}}

\author{J. M. Mercero$^{\aleph}$, M. Rodr\'{\i}guez-Mayorga$^{\dag}$, \\E. Matito$^\ddag$, X. Lopez and J. M. Ugalde}
\date{Kimika Fakultatea, Euskal Herriko Unibertsitatea (UPV/EHU)
and Donostia International Physics Center (DIPC), P.K. 1072, 20080 Donostia,
Euskadi (Spain), $^{\aleph}$IZO-SGI SGIker Kimika Fakultatea, 
Euskal Herriko Unibertsitatea (UPV/EHU),  P.K. 1072, 20080 Donostia,
Euskadi (Spain), 
$^{\dag}$Institut de Qu\'{\i}mica Computacional i Cat\`alisi (IQCC), Departament
de Qu\'{\i}mica, Universitat de Girona, Girona, Catalunya (Spain), 
and $^\ddag$IKERBASQUE Basque Foundation for Science, 
48011 Bilbao, Euskadi (Spain).}

\title{The Electron-pair Density Distribution of the $^{1,3}\Pi_{\rm u}$ Excited States of
H$_{2}$}
\begin{document}

\maketitle

\begin{abstract}
The non-monotonic bahavior of the electron repulsion energy and the 
interelectronic distance, as a function of the internuclear separation,
in the $^{3}\Pi_{u}$ excited state of the hydrogen molecule has been
assessed by explicitly calculation and analysis of the electron-pair density
distribution functions from high level ab initio Full Configuration Interaction wave functions,
for both the $^{3}\Pi_{u}$ and the $^{1}\Pi_{u}$ states.
Additionally, the Hund's rule as applied to these two states has been accounted for
in terms of simple electronic shielding effects induced by wave function 
antisymmetrization.
\end{abstract}

\section{Introduction}

Electron-pair densities describe the relative motion of any
two electrons of the system and were first introduced by Coulson et al. 
to gain insight into the physical consequences of
electron correlation \cite{CN61,CC65,B1/1/10}. Nowadays, however, 
interest on electron-pair densities stems from their usage
to develop faster and more accurate computational methods within both, the
molecular orbitals theory  \cite{KUT-2003} and the density functional theory frameworks \cite{mai:00}.
Additionally, electron-pair densities have recently been used
to unveil the distinctive features of the two-electron
density in different types of chemical bonds 
\cite{piris-2008,Hollet:2011,zielinski:2013,Hennessey:2014,jason-1}.

Electron-pair densities do also reveal, 
even for electronic ground states
\cite{pearson:2009,Per:2009,Wang:2010},
a number of features of the quantum correlations between electrons 
that are challenging to predict at a first sight for, in many cases, 
they are counterintuitive.
Excited states, as expected, exhibit such counterintuitive effects more commonly.
Thus, the double-well first and second excited states of $^{1}\Sigma_{g}^{+}$
symmetry, known respectively as the EF and GK excited states, of the hydrogen molecule
show an intriguing non-monotonic behavior of the mean electron-electron distance
with respect to increasing the internuclear distance. Indeed, at sharp variance
with the ground state \cite{boy:88}, the mean electron-electron distance decreases
as the internuclear distance increases in the transition from
the E to the F minima \cite{wang-0}, and in the transition from the 
G to the K minima \cite{wang-2}, respectively.

%\cite{wang-0,wang-1,wang-2}

In this vein, Tal and Katriel \cite{tal:74} and Colbourn \cite{Colbourn75} reported
the (counterintuitive) non-monotonic behavior of the electron repulsion
energy in the $^{3}\Pi_{u}$ excited state of H$_{2}$.
Indeed, based on their (crude) Hartree-Fock (HF) calculations,
with a small basis set consisting of four uncontracted $sp$
primitives, they found that an increase of the internuclear distance
carries an increase of the electron repulsion energy and a
concomitant decrease of the mean interelectronic distance,
in the domain of the short internuclear distances.
This remarkable counterintuitive feature is not seen
in the parent, as arising from the same $1\sigma^{1}1\pi_{u}^{1}$
configuration, $^{1}\Pi_{u}$ excited state. For this state,
the electron repulsion energy decreases monotonically as the internuclear
separation increases, in the whole range of internuclear separations,
in accordance with common (classical) intuition.
One is naturally prone to attribute this unexpected counterintuitive
behavior of the triplet state to the expected failure of 
HF method for states like these ones which bear substantial
multiconfigurational character, in spite of Tal and Katriel
hypothesis:"{\it $\ldots$the non-monotonic trend
is real rather than a Hartree-Fock artifact"}.

In this paper, electron-pair densities obtained from
high-level ab initio Full Configuration Interaction calculations
will be used to examine these issues
and to put into proper perspective earlier preliminary calculations \cite{mer:03},
demonstrating that the Tal and Katriel hypothesis is true.

\section{Calculations}
The radial electron-pair density distribution, $h(u)$,
of an electronic state $\mid\Psi\rangle$, is:
\begin{equation} \label{bat}
h(u)=u^{2}\int I({\bf u})\, d\Omega_{{\bf u}}
\end{equation}
where, $I({\bf u})$, the so--called \cite{tha87} intracule density, 
\begin{equation} \label{bi}
I({\bf u})=\langle\Psi\mid\sum_{i>j}\delta({\bf u}-{\bf r}_{i}+{\bf r}_{j})\mid\Psi\rangle
\end{equation}
stands for the probability density of the coordinates ${\bf u}-{\bf r}_{i}, \,{\bf r}_{j}$ of
any two electrons to
be separated by the vector ${\bf u}$.
$\Omega_{{\bf u}}$, in Eq. (\ref{bat}), stands for the solid angle subtended the interelectronic
vector ${\bf u}$. 

Observe that the moments of radial electron-pair density,
\begin{equation} \label{hiru}
\langle u^{n} \rangle=\int_{0}^{\infty} u^{n}h(u)\, du
\end{equation}
yield various interesting two--electron properties, like the electron
repulsion for $n$=--1, the number of electron pairs, $n$=0, and the
mean interelectronic distance for $n$=1. Additionally, it is worth noting
that the intracule density can be inferred from accurate total X-ray
intensities \cite{wat:02}.
\begin{figure}[htb]
\epsfig{file=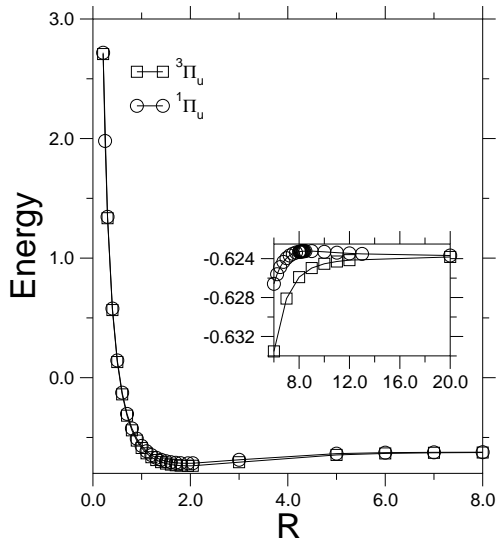,scale=0.4,angle=-90.}
\caption{Calculated potential energy curves for the
$^{1,3}\Pi_{u}$ excited states of H$_{2}$.
Energy and internuclear distance, $R$, in a.u.}
\label{f1}
\end{figure}

\section{Results}
We have calculated the intracule density, $I({\bf u})$, and
its spherically averaged electron-pair density distribution function, $h(u)$,
for both the $^{3}\Pi_{u}$ and the $^{1}\Pi_{u}$ states
of H$_{2}$ from an accurate {\it Full Configuration Interaction} (FCI)
wave function constructed from a large Gaussian basis set
which is described in detail in Ref. \cite{6/5/3}.

The calculated potential energy curves resulting from the calculations
are shown in Figure \ref{f1}, and Table \ref{t2} gives the spectroscopic constants
calculated at the equilibrium geometries, along with the available experimetal data.
Observe that the equilibrium distance of both states and the harmonic vibrational 
frequencies, $\omega_{e}$, 
%and the anharmonicity constants, $\omega_{e}x_{e}$,
are given rather accurately with repect to their experimental marks.

The inset graph of Figure \ref{f1} shows that the $^{1}\Pi_{u}$ state rises
above the dissociation limit asymptote at $R=7.17$ a.u., and reaches
a tiny maximum at the large internuclear distance of $R=9.0$ a.u. Its 
height with respect to the dissociation asymptote is 0.014 eV.
These results are consistent with respect to earlier calculations
of the potential energy curve of this state \cite{C2/22/5}
and lend support to the accuracy of our calculated wave functions.
%The calculated energies of these states are shown in Table \ref{t2},
%and compared with the benchmark results of Rothenberg and
%Davidson \cite{rot:66}. Data shown Table \ref{t2} is very supportive
%of the accuracy of our wave functions.
\begin{table}[htb] 
\begin{center}
\caption{Equilibrium distances, in a.u, Energies, in a.u., vibrational frequencies, in cm$^{-1}$,
%anharmonicity constants, in cm$^{-1}$, 
electron repulsion energies, in a.u., and
electron-electron coalescence densities, in a.u, for the $^{1,3}\Pi_{u}$ excited states of H$_{2}$.
Experimenatal values, in parenthesis, from Ref. \cite{H-1950}.}
\vspace{0.3cm}
\label{t2}
\begin{tabular}{lcc} \hline \vspace{0.1cm}
                       & $^{1}\Pi_{u}$          &  $^{3}\Pi_{u}$ \\ \cline{2-3} \vspace{0.1cm}
$R_{e}$                & 1.95                   & 1.96 \\
                       & (1.952)                & (1.961) \\
$E$                    & --0.716055             & --0.736850 \\
$\omega_{e}$           & 2446.2                 & 2460.9     \\
                       & (2442.7)               & (2465.0) \\
%$\omega_{e}x_{e}$      & 47.83                  & 40.42    \\
%                       & (67.03)                & (61.40) \\
$\left<u^{-1}\right>$  & 0.229863               & 0.246438 \\
$  I(0)$               &  0.81$\times$10$^{-2}$ &  0.26$\times$10$^{-6}$ \\
%
%State    &$R_{e}$&$-E$       &$\omega_{e}$&$\omega_{e}x_{e}$&$<u^{-1}>$  &$I(0)$   \\ \hline
%$^{1}\Pi$&1.95  &0.7160551476&---         & ---             &0.229863194 & 0.81$\times$10$^{-2}$\\
%$^{3}\Pi$&1.96  &0.7368502315&---         & ---             &0.246437290 & 0.26$\times$10$^{-6}$\\
%%1.95   &0.717946 & 0.217651  & 5.67413   & 0.008052   \\
%%       &{\it 0.717965} &   &   &    \\
%%2.05   &0.717458 & 0.215855  & 5.711587  & 0.007728   \\ 
%%       &{\it 0.717439} &   &   &    \\
%%       &         & $^{3}\Pi$ &          &              \\
%%1.95   &0.737260 & 0.241831  & 4.971931  & 0.00\\
%%       &{\it 0.736977} &   &   &    \\
%%2.05   &0.736854 & 0.239319  & 5.013047  & 0.00\\ 
%%       &{\it 0.736455} &   &   &    \\
\hline
\end{tabular}
\end{center}
\end{table}

\begin{figure}[htb]
%\vspace{-2.5cm}
%\begin{center}
\epsfig{file=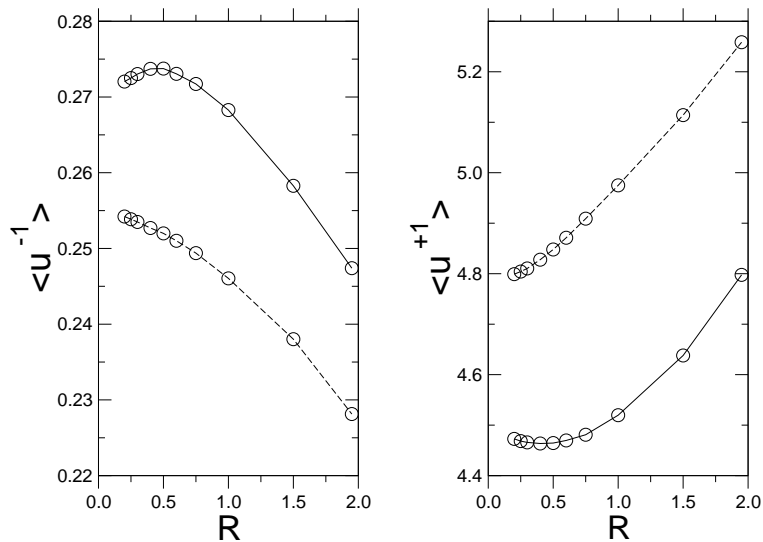,scale=0.4,angle=-90.}
%\epsfig{file=./f2.ps,scale=0.6,angle=0}
%\vspace{-4.5cm}
%\end{center}
\caption{Dependence of the mean interelectronic repulsion energy, $\left<u^{-1}\right>$,
(left panel) and the mean interelectronic separation, $\left<u^{+1}\right>$,
(right panel) in the $^{3}\Pi_{u}$ state
(solid curve) and in the $^{1}\Pi_{u}$ state (dashed curve).}
\label{f2}
\end{figure}

The calculated mean values ($n=\pm1$) of the intracular coordinate $u$,
evaluated as in Eq. (\ref{hiru}), are
shown in Figure \ref{f2} as a function of the internuclear
separation, $R$. 
The counterintuitive bahavior of both the electron repulsion
energy and the interelectronic distance, within the
domain of short internuclear separations, i. e.: $R\in\left[0.2-0.5\right]$ a.u.,
for the $^{3}\Pi_{u}$ state is readily seen upon inspection of
Figure \ref{f2}, which is in sharp contrast with the smoothly
monotonic behavior observed for its parent $^{1}\Pi_{u}$ state.

%The non-monotonic behavior of both the mean
%interelectronic repulsion energy and the mean interelectronic
%separation as a function of $R$ of the $^{3}\Pi_{u}$ state, in
%sharp contrast with the monotonic behavior observed in the
%$^{1}\Pi_{u}$ state, is clearly seen from the inspection of the graphs.
%This confirms that the unusual behavior of the electron-pair density in the
%$^{3}\Pi_{u}$ state of H$_{2}$ is not due the inaccuracy of the
%HF wave function, in accordance with the Tal and Katriel's hypothesis, 
%as one might thought of states like these ones which bear 
%substantial multiconfigurational character, and consequently, their
%HF description is expected to be not reliable.
This puts in place that the non-monotonic behavior of the electron repulsion
and its associated interelectronic distance in the $^{3}\Pi_{u}$ state, in the domain of short
internuclear distances, is not an artifact arising from the crudeness of
its HF description.

\begin{figure}[htb]
%\vspace{-3.0cm}
\epsfig{file=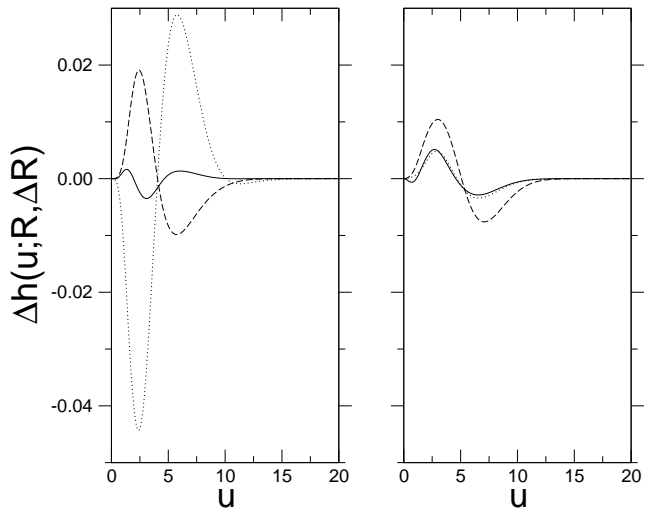,scale=0.4,angle=-90.}
%\epsfxsize\textwidth\epsffile{./f3.ps}
%\vspace{-4.5cm}
\caption{Difference of the electron-pair density probability function
for the $^{3}\Pi_{u}$ state (left panel) and for the
$^{1}\Pi_{u}$ state (right panel). Solid curve:
$h(u;R=0.2)-h(u;R=0.5)$, dotted curved: $h(u;R=0.5)-h(u;R=0.75)$,  
and dashed curve: $h(u;R=1.5)-h(u;R=1.95)$}
\label{f3}
\end{figure}

Inspection of the difference between the electron-pair
density distribution functions calculated at two internuclear
distances, namely,
\begin{equation}
\Delta h(u;R,\Delta R)= h(u;R)- h(u;R+\Delta R),\; \Delta R>0
\end{equation}
provides an alternative view 
of these unusual correlation effects, 
relative to the more familar $h(u)-h_{HF}(u)$ difference.
Indeed, as seen in Figure \ref{f3}, we observed that for the
$^{1}\Pi_{u}$ state, increasing the internuclear distance from
$R=0.2$ a.u. to $R=0.5$ a.u., from
$R=0.5$ a.u. to $R=0.75$ a.u. and from $R=1.5$ a.u to $R=1.95$ a.u.
results in a decreased probability of finding the electrons at short distances and
a concomitant increased probability of finding the electrons at
larger distances. Notice that the three curves of the right panel
of Figure \ref{f3} are positive for small interelectronic distances,
hence the probability of finding two electrons
within these short interelectronic distances is larger for the small 
internuclear distance, and vice-versa for large interelectronic distances.

However, for the $^{3}\Pi_{u}$ state,
the probability of finding the electrons at short relative distances is 
larger for  $R=0.2$ a.u. than for $R=0.5$ a.u, in spite of the tiny
positive peak at $u\sim1.25$ a.u., and clearly much larger for
$R=0.75$ a.u. than for $R=0.5$ a.u. (see dotted curve of the left panel of Figure \ref{f3}),
opposite to what is found for larger internuclear distances. For instance, 
the probability of finding the two electrons close to each other is larger at
$R=1.5$ a.u. than at $R=1.95$ a.u., in accordance with intuition.

But, as mentioned above, at smaller internuclear distances,
increasing the internuclear distance, 
%from $R=0.5$ a.u. to $R=0.75$ a.u. 
increases the probability of finding the electrons at short 
interelectronic distances. This behavior is {\it counterintuitive},
and should be seen as one more ({\it unexpected}) effect of the 
symmetry constraints imposed by the Pauli principle.

\section{Hund's Rule in the $^{1,3}\Pi_{u}$ States of H$_{2}$.}

%The main difference between the $^{1,3}\Pi_{u}$ states arises from
%the symmetry of the spatial part of the wave function. Since,
%the triplet spin wave function is symmetric, its associated
%spatial part must be antisymmetry under exchange of
%electrons, whereas for the singlet the
%antisymmetry of the spin wave function dictates that the
%spatial part must be symmetric under exchange of electrons.
%One immediate consequence of this is that the intracular
%density must be zero at ${\bf u}$=0 for the triplet state.
%Recall that the for an antisymmetric function, like the spatial
%part of the triplet state,
The parent  $^{1,3}\Pi_{u}$ states of the hydrogen molecule differ
because of the different symmetry constraints which Pauli's principle
imposes to the spatial part of their corresponding wave functions. Thus,
while the singlet state transforms symmetrically with respect to 
exchanging the electronic coordinates, ${\bf r}_{1}\leftrightarrow{\bf r}_{2}$,
the triplet state's spatial part of the wave function must do it
antisymmetrically, namely,
\begin{eqnarray}
\label{e30}
\Psi({\bf r}_{1},{\bf r}_{2})=-\Psi({\bf r}_{2},{\bf r}_{1}),\;\;\forall({\bf r}_{1},{\bf r}_{2})
\end{eqnarray}
and, consequently,
\begin{eqnarray}
\label{e31}
\Psi({\bf r}_{1},{\bf r}_{1})=0,\;\;\forall{\bf r}_{1}
\end{eqnarray}
This allows for the straightforward evaluation of the electron-electron coalescence
density \cite{US94}, namely: $I({\bf u}=0)$, for the triplet state as:
%Therefore, in accordance with Eq. (\ref{bi}),
\begin{eqnarray}
\label{e32}
I(0)&=&\langle\Psi\left|\delta({\bf r}_{1}-{\bf r}_{2})\right|\Psi\rangle
\nonumber \\ &=&
\int d{\bf r}_{1}d{\bf r}_{2}\Psi^{\ast}({\bf r}_{1},{\bf r}_{2})
\Psi({\bf r}_{1},{\bf r}_{2})\delta({\bf r}_{1}-{\bf r}_{2})
\nonumber \\ &=&
\int d{\bf r}_{1}\Psi^{\ast}({\bf r}_{1},{\bf r}_{1})
\Psi({\bf r}_{1},{\bf r}_{1})=0
\end{eqnarray} 
Our explicitly calculated values of $I(0)$ for the $^{3}\Pi_{u}$ state,
shown in Table \ref{t2} agree with this prediction,
and lend further support to our calculated intracule densities.

Furthermore, due to the continuity of the intracule density function,
it is expected that the spherically averaged electron-pair density distribution
function, $h(u)$,
will start building up slower in the triplet state than in the singlet,
because in the singlet state $I(0)>0$ (see Table \ref{t2}). 
Consequently one expects that the probability of finding two electrons at short 
interelectronic distances will be larger for the singlet than for the triplet. 

\begin{figure}[htb]
\epsfig{file=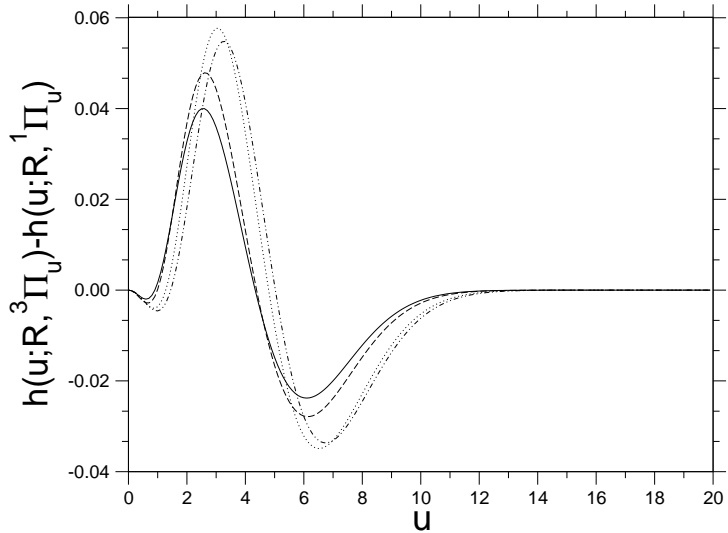,scale=0.4,angle=-90.}
\caption{Difference between the electron-pair density probability functions
of the $^{3}\Pi_{u}$ state and the $^{1}\Pi_{u}$ state. Solid curve:
$R=0.2$ a.u. Dashed curve $R=0.5$ a.u. Dotted curve: $R=1.5$ a.u.
Dotted and dashed curve: $R=1.95$ a.u.}
\label{f4}
\end{figure}

The electron-pair density distribution function differences of the 
$^{3}\Pi_{u}$ state minus that of $^{1}\Pi_{u}$ state, 
at a number of selected internuclear distances,
plotted in Figure \ref{f4}, confirm this assumption. 
Namely, as stated above $h(u)$ is smaller at small interelectronic
distances, $u$, for the triplet than for the singlet, hence the
negative values shown in Figure \ref{f4} at short interelectronic
distances $u$, irrespective of the internuclear distance.

%This has been argued to constitute the physical basis of the
%Hund's rule \cite{HUND,HUND-1,HUND-2}. Namely, the lower probability of finding
%electron pairs close together in the triplet with respect to
%the singlet results in a greater electron repulsion
%in the low spin state as compared to the high spin state and, consequently,
%the high spin state should be more stable. Numerous 
%calculations on a variety of systems (see page 234 of Ref.\cite{VUB00}
%for a full list of references) and our own
%ones reported in Table \ref{t2} show, however, that
%this explanation is incorrect \cite{Russ-1,KP1977,Russ-2,WB1985}. The electron
%repulsion energy is indeed larger for the $^{3}\Pi_{u}$ state that for the
%$^{1}\Pi_{u}$ state. The physical basis for the larger of
%stability of the triplet with respect to the singlet is,
%nevertheless, insinuated in Figure \ref{f4}.
The Pauli principle, therefore, prevents electrons to come
into close proximity of each other, as is well known. 
A natural consequence of this is (hypothesized) that the electron
repulsion in the triplet state should be smaller than in its parent
same-configuration singlet state, where electrons are not
impeded to approach each other, and consequently, due to
associated decreased electron repulsion energy, the triplet (high)
spin state results to be more stable that the singlet (low) spin state.
This has been claimed to constitute the physical basis of the
Hund's rule \cite{HUND,HUND-1,HUND-2}, which the $^{1,3}\Pi_{u}$ excited states of H$_{2}$ 
strictly fulfill.

However, the data reported in Table \ref{t2} and in Figure \ref{f2}, shows that
this is not the case for the $^{1,3}\Pi_{u}$ excited states of H$_{2}$.
Indeed, the electron repulsion energy for the triplet state is larger than for the
singlet state, irrespective of the internuclear distance. 
Additionally, it is worth recalling that numerous explicit evaluations
of the electron repulsion energy for the various spin states arising
from the same configuration found, with no exception, that the electron
repulsion energy is larger in the high-spin state (see Ref. \cite{VUB00}, page 234).
This invalidates the explanation outlined above for the 
Hund's rule as it was elegantly put forward by Boyd \cite{Russ-1,Russ-2}
and subsequently elaborated by others \cite{KP1977,WB1985}.

The physical basis of the lower energy of the $^{3}\Pi_{u}$,
with respect to its parent $^{1}\Pi_{u}$ state, is drawn in Figure \ref{f4}.
Notice that although the probability of finding the electron in close proximity
is smaller in the triplet than in the singlet,
the triplet favors intermediate interelectronic
distances, as compared to the singlet state.
Additionally, notice also that the probability of finding the electrons 
at large separation, is larger in the singlet
than in the triplet, alike the behavior found for short interelectronic distances. 
The triplet state, therefore, favors intermediate interelectronic distances 
which makes the electronic cloud more compact in the triplet than in the singlet and,
consequently, makes the electron-nucleus attraction energy
larger in the triplet than in the singlet, in such an amount that it outweighs
the larger electron repulsion of the latter
\cite{kat:72,Regier84a,uga:85,Thakkar87e,dar:87}.

In other words, since the electrons of the triplet avoid 
each other in the vicinity of the nuclei they
screen less the nuclear charge and, consequently the
electron cloud gets more compact than in the singlet
for which the nuclear charge is screened more efficiently \cite{WBB1980}.
This leads ultimately to an increased electron-nucleus
attraction for the triplet, which overweights the larger electron repulsion
of the triplet yielding, therefore, a more stable triplet state.

\section{Summary}

We have demonstrated, in accordance with Tal and Katriel \cite{tal:74},
that the non-monotonic behavior with respect to the
internuclear separation of the electron repulsion energy and its
associated mean interelectronic distance in the $^{3}\Pi_{u}$ excited state
of the hydrogen atoms are real, {\it counterintuitive}, effects
of the symmetry constraints imposed by the Pauli principle on the wave function of triplet
states. High-level Full Configuration Interaction calculations
show that while in the $^{1}\Pi_{u}$ excited state
the electron repulsion energy and its
associated mean interelectronic distance
behave monotonically, in the $^{3}\Pi_{u}$ excited state,
the electron repulsion energy increases and the 
mean interelectronic distance decreases
as the internuclear separation increases.

Finally, we have found that the Hund's rule, which holds also
for these $^{1,3}\Pi_{u}$ same-configuration, $1\sigma^{1}_{g}\pi^{1}_{u}$, 
excited states, can be accounted for
in terms of simple electronic shielding effects induced by wave function 
antisymmetrization, in consonance with the accepted interpretation \cite{Russ-2}.

\section{Acknowledgment}
This research has been funded by Euskal Herriko Unibertsitatea
(the University of the Basque Country),
Eusko Jaurlaritza (the Basque Government) and the Spanish
Office of Science and Technology (MINECO CTQ2014-52525-P).
JMU wishes to thank Prof. Russell J. Boyd for his continuous encouragement
and dedicated guidance over the years across the exciting field of 
electron-pair densities.

%\bibliographystyle{unsrt}
%\bibliography{nato,h2}

\end{document}